\documentstyle[11pt,paspconf]{article}
\def\etal{{\it et al.\ }}

\def\eg{{\it e.g.,\ }}

\def\ie{{\it i.e.,\ }}
\def\msol{\ifmmode {\>M_\odot}\else {$M_\odot$}\fi}
\def\cmsq{\ifmmode {\>{\rm\ cm}^2}\else {cm$^2$}\fi}
\def\psqcm{\ifmmode {\>{\rm cm}^{-2}}\else {cm$^{-2}$}\fi}
\def\psqpc{\ifmmode {\>{\rm pc}^{-2}}\else {pc$^{-2}$}\fi}
\def\pcsq{\ifmmode {\>{\rm\ pc}^2}\else {pc$^2$}\fi}
\def\intensity{\ifmmode{{\rm erg\ cm}^{-2}{\rm\ s}^{-1}
      {\rm\ Hz}^{-1}{\rm\ sr}^{-1}}
      \else {erg cm$^{-2}$ s$^{-1}$ Hz$^{-1}$ sr$^{-1}$}\fi}
\def\iintensity{\ifmmode{{\rm erg\ cm}^{-2}{\rm\ s}^{-1} {\rm\
sr}^{-1}}\else
      {erg cm$^{-2}$ s$^{-1}$ sr$^{-1}$}\fi}
\def\flux{\ifmmode{{\rm erg\ cm}^{-2}{\rm\ s}^{-1}}\else {erg
cm$^{-2}$ s$^{-1}$}\fi}
\def\fluxdensity{\ifmmode{{\rm erg\ cm^{-2}\ s^{-1}\ Hz^{-1}}}\else {erg
cm$^{-2}$ s$^{-1}$ Hz$^{-1}$}\fi}
\def\phoflux{\ifmmode{{\rm phot\ cm}^{-2}{\rm\ s}^{-1}}\else {phot
cm$^{-2}$ s$^{-1}$}\fi}
\def\pintensity{\ifmmode{{\rm phot\ cm}^{-2}{\rm\ s}^{-1}
      {\rm\ Hz}^{-1}{\rm\ sr}^{-1}}
      \else {phot cm$^{-2}$ s$^{-1}$ Hz$^{-1}$ sr$^{-1}$}\fi}
\def\epuv{\ifmmode{{\rm erg\ cm}^{-3}{\rm\ s}^{-1}}\else {erg
cm$^{-3}$ s$^{-1}$}\fi}
\def\lum{\ifmmode{{\rm erg\ s}^{-1}}\else {erg s$^{-1}$}\fi}
\def\pholum{\ifmmode{{\rm phot\ s}^{-1}}\else {phot s$^{-1}$}\fi}
\def\nt{\ifmmode{{\rm cm^{-3}\ K}}\else {cm$^{-3}$ K}\fi}
\def\be{\begin{equation}}
\def\ee{\end{equation}}
\def\bea{\begin{eqnarray}}
\def\eea{\end{eqnarray}}
\def\beas{\begin{eqnarray*}}
\def\eeas{\end{eqnarray*}}
\def\Em{\ifmmode{{\cal E}_m}\else {{\cal E}$_m$}\fi}
\def\Tkev{\ifmmode{T_{\rm kev}}\else {$T_{\rm keV}$}\fi}
\def\emunit{\ifmmode{{\rm cm}^{-6}{\rm\ pc}}\else {cm$^{-6}$ pc}\fi}
\def\fesc{\ifmmode{f_{\rm esc}}\else {$f_{\rm esc}$}\fi}
\def\hubunits{\ifmmode {\>{\rm km\ s^{-1}\ Mpc^{-1}}}\else {km
s$^{-1}$ Mpc$^{-1}$}\fi}
\def\gta{\;\lower 0.5ex\hbox{$\buildrel > \over \sim\ $}}
\def\lta{\;\lower 0.5ex\hbox{$\buildrel < \over \sim\ $}}          
\def\kms{\ifmmode {\>{\rm km\ s}^{-1}}\else {km s$^{-1}$}\fi}

\begin{document}

\title{Ionizing Photon Sources within the Local Group}

\author{Philip R. Maloney}
\affil{Center for Astrophysics and Space Astronomy, University of Colorado}
\author{J.~Bland-Hawthorn}
\affil{Anglo-Australian Observatory}

\begin{abstract}
We review the possible sources of ionizing photons within the Local
Group. Throughout most of the LG volume, the cosmic background
radiation will dominate, but locally (e.g., within $d\sim 100$ kpc of
the Galactic disk) stellar ionizing photons escaping from galaxies may
dominate. The magnitude of the cosmic ionizing background should be
determined in the very near future by observations of H$\alpha$
emission from the outer neutral hydrogen disks of late-type
spirals. The detection of the Magellanic Stream in H$\alpha$ suggests
that a few percent of the ionizing photons produced in the Galaxy
escape the disk. Neither a warm Local Group corona nor decay photons
from a neutrino halo can explain the Stream emission.
\end{abstract}


\section{Introduction}
The magnitude and origin of the flux of ionizing photons within the
Local Group of galaxies is of considerable interest, for two reasons:
\begin{enumerate}
\item Although the total ionizing photon luminosity of the Galaxy is
fairly well known, the details of the transfer of this radiation and,
in particular, what fraction of the ionizing photons escape to large
vertical heights above the midplane - or even escape the disk entirely
- has been very uncertain. A significant escaping flux has major
implications for many aspects of the interstellar medium.
\item Determination of the level of any cosmic background of ionizing
radiation has important cosmological ramifications, since this
background (which {\it must} exist at some level) arises in a known
population of objects, either active galactic nuclei (AGN) or
star-forming galaxies, or a combination of the two.
\end{enumerate}
Here we review the possible sources of ionizing photons within the
Local Group and the current constraints. After a brief discussion of
terminology, we review first the possible cosmological sources, and
the present constraints on the magnitude of a cosmic background. We
then discuss local (\ie within the Local Group) sources, for at least
one of which (ionizing photons escaping from the Milky Way) we now
have a direct observational determination. In the final section we
discuss the implications of these results, for both the interstellar
medium (and, in particular, high-velocity clouds) and cosmology. 

\section{Terminology}
The magnitude of the ionizing background at large redshift is
extremely important for the structure of the intergalactic medium, as
it sets the level of ionization in the IGM, and there has been a
considerable amount of effort to determine the intensity of the
ionizing background. The standard quantity which has been generally
used in these studies is $J_o$, the mean intensity (in \intensity) at
the Lyman limit. This is rather unfortunate, because for comparison
with observations at $z=0$ (where, as discussed below, the limits on
the ionizing background come from observations of recombination
lines), the most useful quantity is the normally incident ionizing
photon flux, 
\bea \phi_i&\equiv& \int_0^1 \mu\,d\mu \int\limits_{\nu_o}^\infty 
{4\pi J_\nu\over h\nu}\,d\nu \\
&=&4.73\times 10^3 {J_{-23}\over\alpha}\;\phoflux 
\eea 
where $\mu$ is
the cosine of the angle with respect to the normal, $J_o=10^{-23}
J_{-23}$ \intensity\ and the spectral index, $\alpha$, is defined so
that the monochromatic flux $f_\nu\propto \nu^{-\alpha}$. This is
simply related to the emission measure, 
\be 
\Em\equiv \int n_e n_i\,dl 
\ee 
(where $n_e$ and $n_i$ are the volume densities of electrons
and ions, and $dl$ is an element of path length along the line of
sight) by 
\be 
\Em=2.5\times 10^{-2}\phi_4\;\emunit 
\ee
 where $\phi_i=10^4\phi_4\;\phoflux$ and the numerical coefficient
assumes a gas temperature $T=10^4$ K and a face-on slab that is
optically thick to the ionizing photons and is being irradiated from
both sides. This latter assumption is true for the cosmic background,
but will not in general be correct for local sources (\eg photons
escaping from the Galactic disk), where the irradiation is
one-sided. The corresponding H$\alpha$ surface brightness (again for
two-sided illumination) is $I_{\rm H\alpha}=9\phi_4$ mR, where one
milliRayleigh is $10^3/\pi$ \phoflux\ sr$^{-1}$. The H$\alpha$ flux
for a given $\phi_i$ is essentially independent of both density and
temperature, since raising or lowering the recombination rate simply
alters the emission measure proportionately, leaving the total column
recombination rate unaltered.

\section{Sources: Cosmological}
Possible cosmological sources fall into two categories: standard (AGN
and stellar ionizing photons from galaxies) and exotic (decaying
particles). At present the only possibility which is worth discussing
in the latter category is Sciama's decaying neutrino model; as we will
see below, this also counts as a local source of ionizing photons. 

\subsection{Standard} Estimates of the contribution of QSOs to the
ionizing background as a function of redshift have a long history (\eg
Sargent \etal 1979; Bechtold \etal 1987; Miralde-Escud\'e \& Ostriker
1990; Haardt \& Madau 1996). The most recent values have declined by
about a factor of four from the earliest estimates; Haardt \& Madau
(1996) find $J_{-23}\sim 1$ at $z=0$. Shull \etal (1998) have included
the contribution from Seyfert galaxies as well as QSOs, and find a
slightly larger value: $J_{-23}\sim 1.6$. For a canonical spectral
index at the relevant photon energies of $\alpha\approx 1.8$ (Zheng
\etal 1997), this implies a photon flux $\phi_i=4200\;\phoflux$.  It
must be noted that these estimates depend not only on the adopted
luminosity functions, but also on the magnitude of filtering of the
AGN spectrum by the intergalactic medium, which depends on both the
spectral shape and the structure which is adopted for the IGM (\eg
Fardal, Giroux, \& Shull 1998).

In some ways estimating the ionizing background due to galaxies is
more straightforward, since the galaxy luminosity function at low
redshift is fairly well understood, and the high ionization of the IGM
at the present-day minimizes its influence on the propagation of
ionizing photons. (There are, however, still significant uncertainties
in the ionizing photon luminosities and the continuum shape for hot
stars: see Schaerer \& de Koter 1997, and references therein.)
Giallongo \etal (1997) estimate $J_{-23}\sim 130 \fesc$ at low
redshift ($z\approx 0.5$). It is here that the major uncertainty in
evaluating the galactic (\ie stellar) contribution to the ionizing
background enters: what fraction \fesc\ of the ionizing photons
produced by hot stars within a galactic disk escape into the IGM?
Theoretical models of the transport of ionizing radiation within the
Galactic disk suggest that approximately 10\% of the ionizing photons
produced within the Milky Way escape the disk entirely (Dove \& Shull
1994a). However, observational determinations are clearly
necessary. 

From observations of four UV-bright starburst galaxies with
the {\it Hopkins Ultraviolet Telescope}, Leitherer \etal (1995)
determined upper limits to \fesc\ of 1.7, 4.8, 15, and 1\%, (for IRAS
08339+6517, Mrk 1267, Mrk 66, and Mrk 496, respectively) and concluded
that the escaping fraction must be small. Their analysis did not take
into account the absorption of ionizing radiation by the ISM of the
Milky Way, however, and as shown by Hurwitz, Jelinsky \& Dixon (1997),
this correction raises the upper limits listed above to 5.2, 11, 57,
and 3.2\%, so that they are no longer significant constraints. (Note
also that Hurwitz \etal did not include the effects of absorption by
molecular hydrogen, which could raise these upper limits further; see
the discussion in \S 4 of their paper.) As we will see below, recent
observations of H$\alpha$ emission from the Magellanic Stream suggest
that for the Galaxy, $\fesc\approx 6\%$. 

\subsection{Exotic} In a long series of papers (\eg Sciama 1990,
1991, 1995a,b), Sciama has argued that a wide range of ionization
problems in both the interstellar and intergalactic medium can be
understood as the consequence of the decay of a massive neutrino
species (presumably the $\tau$ neutrino) to a less massive species,
accompanied by emission of a photon with energy $E > 13.6$ eV. (Decay
occurs if the neutrino weak (\ie flavor) eigenstates are not also mass
eigenstates; see \eg Boehm \& Vogel 1992.)  This model predicts a
cosmic background due to the decaying neutrinos, with magnitude set by
the decay lifetime, $T_\nu$, and by the neutrino mass, $m_\nu$, which
determines the photon energy. If one assumes that the mass of the
decay product is $\ll$ the mass of the decaying neutrino, then the
photon energy produced by the decay is simply $E\simeq m_\nu
c^2/2$. The most recent values of the parameters are summarized in
Sciama (1998):
\beas
m_\nu&\simeq&27.4\;{\rm eV} \\ 
\tau_\nu&=&\left(2\pm 1\right)\times 10^{23}\;{\rm s} \\
\phi_i&<&10^5\;\phoflux 
\eeas
where the upper limit to the cosmic ionizing flux is (by construction)
of the order of current upper limits (as measured by H$\alpha$
observations; see below). If the decay photon energy is close to the
hydrogen ionization threshold, then the cosmic ionizing background is
limited by the redshifting of the photon energy with increasing $z$;
this places stringent constraints on the mass of the decaying
neutrino. Under the assumption that the mass of the universe is
dominated by massive neutrinos (\ie $\Omega_\nu \simeq 1$, with one
flavor of neutrino much more massive than the others), the Hubble
constant is also tightly constrained: $H_o= 55\pm 0.5\;
\hubunits$. Recent H$\alpha$ observations of the late-type spiral
galaxy NGC 3198 (Bland-Hawthorn 1998; Bland-Hawthorn, Veilleux, \&
Carignan 1998, in preparation) are, however, in marked disagreement
with the predictions of Sciama (1995a): the observed emission is much
fainter (by about an order of magnitude) than predicted by the
decaying neutrino model (see \S 4.2 below).

\section{The Cosmic Background: Observations}
\subsection{Indirect}
The most straightforward and model-independent way to determine the
cosmic ionizing background is simply to measure it directly through
observations of recombination lines, such as H$\alpha$, as discussed
in the following section. However, it is only very recently that the
sensitivity of these measurements have begin to reach the necessary
levels. Other, more indirect methods have therefore been used to
estimate the ionizing flux.

The first method is based on the observation that the neutral hydrogen
disks in galaxies appear to have rather abrupt truncations, and, more
generally, there appears to be very little atomic hydrogen at the
boundaries of or outside of galaxies at column densities $N\lta
10^{19}$ \psqcm\ (see summary in van Gorkom 1993). This can be
understood as the consequence of ionization of the galactic hydrogen
disk by the extragalactic radiation field: in a sort of ``inverse
Str\"omgren'' problem, the hydrogen disk will be largely ionized once
the total column drops below a critical value $N_c$, given by
\be
N_c\approx 1.7\times 10^{19}\left[\phi_4 \left({\sigma_{\rm zz}\over
6\;{\rm km\;s^{-1}}}\right)\left({V_{100}\over\Sigma_{100}}\right)
\right]^{1/2}\;\psqcm
\ee
(Maloney 1993), where $\sigma_{\rm zz}$ is the vertical velocity
dispersion of the gas, $V=100 V_{100}\;\kms$ is the galactic rotation
velocity, and $\Sigma=100\Sigma_{100}\;\msol\psqpc$ is the total
surface density (stellar disk + gas + halo; in general, this is
dominated by the halo at the relevant radii). Detailed modelling of
NGC 3198, as shown in Figure 1, suggests that $\phi_4\sim 1$. (Note
also that while this idea was rediscovered following the famous
unpublished NGC 3198 experiment, it actually dates back to Sunyaev
1969 and Bochkarev \& Sunyaev 1977; see Maloney 1993 for a
summary of the history.) The actual value of $\phi_i$ which best
explains the data is somewhat fuzzy, because at gas column densities
near the critical value small variations in $N_{\rm H}^{\rm tot}$ lead
to substantial changes in $N_{\rm HI}$. This is immediately apparent
from Figure 1, in which the neutral column densities for the two
halves of the major axis differ rather dramatically beyond $R\sim 25$
kpc, due to modest variations in $N_{\rm H}^{\rm tot}(R)$. The total
hydrogen column as a function of radius has been ``reverse-engineered'' 
from the observed $N_{\rm HI}(R)$ by inverting the neutral hydrogen
column (for a choice of $\sigma_{\rm zz}$ and $\phi_i$). A difficult
question to address {\it a priori} is how much deviation from a
smoothly varying $N_{\rm H}^{\rm tot}(R)$ (especially at or near the
HI ``edge'') is acceptable in the context of a photoionization
explanation for the HI truncations: as the ionizing background
intensity is increased, the inferred neutral hydrogen column density
becomes smoother. For this reason, the estimates of the necessary
value of $\phi_i$ range about a factor of three to either side of
$\phi_4 = 1$ (Maloney 1993; Corbelli \& Salpeter 1993; Dove \& Shull
1994b). This ambiguity can be eliminated by high-sensitivity H$\alpha$
observations, since a corollary of the photoionization modelling is a
prediction of the emission measure as a function of radius, as
discussed in \S 4.2.

\begin{figure}[!htb]
\plotfiddle{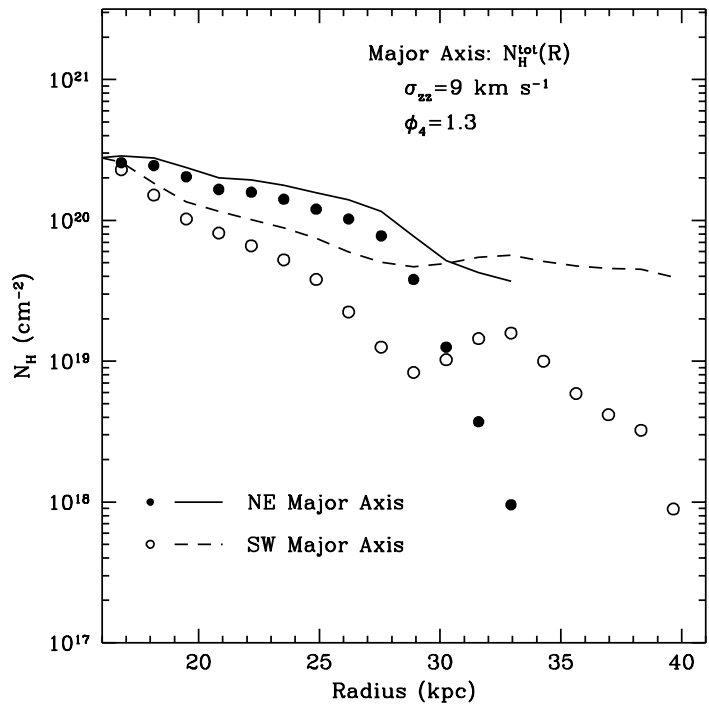}{88truemm}{0}{65}{65}{-200}{-160}
\caption{Observed $N_{\rm HI}$ and derived $N_{\rm H}^{\rm tot}$ for
the NE ({\it filled circles, solid line}) and SW ({\it open circles,
dotted line}) halves of the major axis of NGC 3198. The total hydrogen
column densities are calculated requiring a match to the observed
neutral hydrogen columns to within $0.2\%$.}
\end{figure}

The other indirect method which has been used to estimate the
strength of the cosmic ionizing background is the ``proximity
effect''. The basic idea is simple: in the vicinity of a luminous
source of ionizing photons, such as a quasar, the observed number of
Ly$\alpha$ forest lines will decrease, as the (local) ionizing photon
flux will reduce the neutral fraction in the absorbers near the
source, with the result that for some absorbers the neutral columns
will drop below the threshold for detection as a result of the
enhanced ionization. It is possible to use this to infer the strength
of the cosmic ionizing background (\ie the sum over all sources of
radiation) since the relative importance of the local ionization by a
given quasar is inversely proportional to the background intensity.
This effect was first noticed by Carswell \etal (1982), and the
theoretical analysis was pioneered by Bajtlik, Duncan, \& Ostriker
(1988). Most subsequent studies have been confined to high redshift
($z\approx 3$). However, Kulkarni \& Fall (1993) presented a tentative
detection of the proximity effect in a low redshift sample (with
$z\lta 1$), and derived a mean intensity $J_{-23}\sim 0.6$ at
$z\approx 5$. This value is surprisingly low. It must be kept in mind,
however, that, due to the small sample size, the error bars on this
determination are very large (a point emphasized by Kulkarni \& Fall):
at the $1\sigma$ level, the flux could be larger by a factor of
six. At the moment this method must be regarded only as suggestive,
until much more data are available.

\subsection{Direct} Preferable to any indirect determination of the
background intensity is, of course, a direct one, and this can in
principle be provided by hydrogen recombination line observations. As
outlined in \S 2, provided that the neutral column density in some
region (the outer parts of a galactic disk, or a cloud) is large
enough that it is opaque to the bulk of the incident ionizing photons,
the expected emission measure is independent of the density (and
density structure) within the region, since it is basically being used
as a photon counter: every incident photon is absorbed, and this must
be balanced (in steady-state) by a recombination. The observable
emission measure (\ie as observed in recombination lines) will also be
insensitive to the gas temperature, as noted above: increasing the
recombination rate by reducing the temperature, for example, will
simply reduce the column emission measure by the same factor. The
primary uncertainty in using this method to measure the ionizing
background is the effects of geometry: the observed emission measure
will depend on both the orientation of the observed region with
respect to our line of sight, and on the (areal) filling factor of the
region, as seen by the ionizing photons.

There is a fairly long history of attempts to measure the cosmic
ionizing background using observations of the H$\alpha$ recombination
line (Reynolds \etal 1986; Songaila, Bryant, \& Cowie 1989; Kutyrev \&
Reynolds 1989; Donahue, Aldering \& Stocke 1995; Vogel \etal
1995). This has been hampered by the difficulties of observing such
faint levels of emission. The current best {\it published} upper limit
to the cosmic background is that of Vogel \etal (1995), which is
$\phi_4 < 4.5$ at the $3\sigma$ level. 

An example of what sort of flux levels are necessary is shown in
Figure 2. This is a plot of the predicted emission measure for the
same photoionization model of NGC 3198 shown in Figure 1. It is very
important to note that the values of $\Em (R)$ plotted in Figure 2
have been corrected for disk inclination $i$ (which was {\it not} the
case for the corresponding Figure 14 in Maloney 1993), using $i(R)$
from Begeman (1989). Since the inclination $i\gta 71^\circ$ for $R >
16$ kpc, the resulting increase in path length raises the observable
emission measures by at least a factor of three above the face-on
value for all radii of interest.
\begin{figure}[!htb]
\plotfiddle{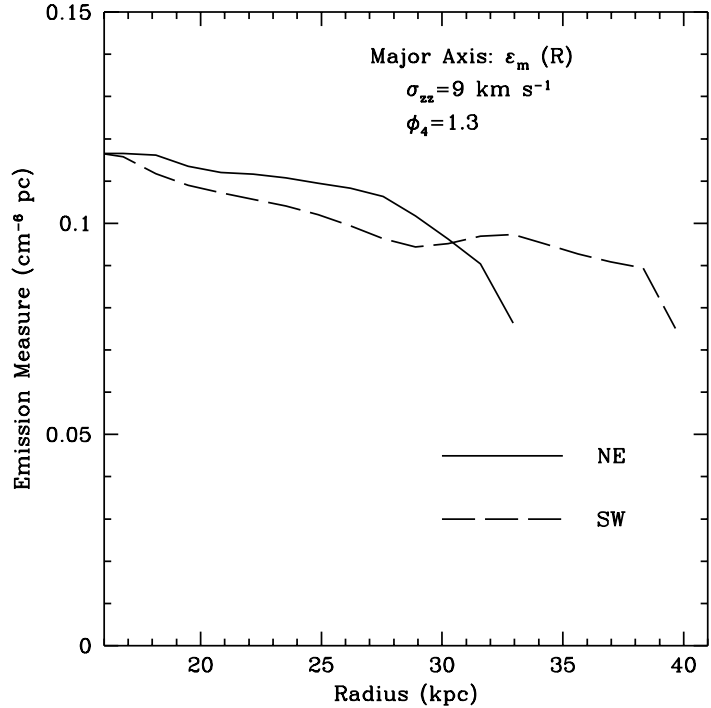}{88truemm}{0}{65}{65}{-200}{-160}
\caption{Predicted emission measures for the NE ({\it solid line}) and
SW ({\it dotted line}) halves of the major axis of NGC 3198. The
values of $\Em (R)$ are those predicted by the photoionization model
shown in Figure 1, and have been corrected for inclination.}
\end{figure}

The data on NGC 3198 presented by Bland-Hawthorn (1998) (also
Bland-Hawthorn, Veilleux, \& Carignan 1998, in preparation), showing
H$\alpha$ emission at the level of $0.1\emunit$, strongly
suggest to the first author that the cosmic background has been
detected, at a level $\phi_4\sim 1$. The second author is less
convinced, and believes the detected emission may be due to internal
sources. There should be two ways of discriminating between these
possibilities:
\begin{itemize}
\item If the emission is really due to the cosmic background, then the
outskirts of other spiral galaxies will show emission at identical
(inclination-corrected) levels. For example, the outer HI disk of M31,
which is at an inclination of about $77^\circ$, should exhibit the
same H$\alpha$ surface brightness as NGC 3198.
\item The radial distribution of emission measure will be crucial. As
is apparent in Figure 2, $\Em(R)$ is predicted to be quite flat in the
case of a uniform cosmic background. This is simply the result of the
fact that nearly all of the incident ionizing photons are absorbed by
the gas disk (see \eg Vogel \etal 1995); thus, until the HI edge is
reached, there is merely a slow radial fall-off as the absorbed
fraction declines slightly with the decrease in the total column
density. Beyond the HI edge the emission measure decreases rapidly,
since most of the gas is now ionized rather than neutral. In contrast,
if the observed emission measure is due to internal sources (such as
self-irradiation of a warped disk), in general \Em\ will be a strong
function of radius, depending on how the irradiation varies with
$R$. Note that \Em\ may either {\it decrease} or {\it increase} with
$R$ in this case; a fuller discussion of the behavior expected in
different scenarios will be presented in Maloney, Bland-Hawthorn \&
Freeman (1998). 
\end{itemize}
Further observations are clearly required to settle this issue;
fortunately, the necessary instrumental sensitivity and data reduction
techniques are in hand, and rapid progress in this area is to be
expected.

\section{Sources: Local}
As for the cosmological background, sources here come in two flavors:
standard, and exotic. Under ``standard'' are ionizing photons produced
by hot stars within Local Group galaxies which escape from the
galactic disks; as we discuss below we now have a good estimate of the
escape fraction for the Milky Way. In addition, there is the
possibility that the ionizing photon flux from a Local Group
``corona'' of warm gas might be significant (Maloney \& Bland-Hawthorn
1998). Finally, Sciama's decaying neutrino model also enters as a
local source, because the neutrinos which make up the dark matter
haloes of the Local Group galaxies (in this model) will make a local
contribution to the ionizing photon flux, in addition to the
cosmological contribution from neutrino decay.
\subsection{Stellar $\phi_i$ from Local Group Galaxies}
The ionizing photon luminosity of the Milky Way is fairly
well-determined; various estimates agree on $N_i\approx 3\times
10^{53}$ \pholum\ (see Bland-Hawthorn \& Maloney 1998), with an
uncertainty of about $30\%$. As noted earlier, the question as to what
fraction of these photons escape from the disk to the halo and beyond
has remained unsettled. However, a recent observation has settled this
question. Weiner \& Williams (1996) detected H$\alpha$ emission from
the Magellanic Stream, with emission measures $\Em\approx 0.5-1$
\emunit. From equation (4) (but corrected to one-sided irradiation),
this implies normally incident ionizing photon fluxes $\phi_i\approx
4-8\times 10^5$ \phoflux. Although Weiner \& Williams argued for a
shock origin, it is difficult to make these models work, and it is far
more likely that the H$\alpha$ emission is produced by ionizing
photons produced in the disk of the Milky Way which have escaped the
disk to impinge upon the Stream. The geometry is extremely favorable,
as the Stream passes more or less directly over the Galactic disk. As
discussed in detail in Bland-Hawthorn \& Maloney (1997, 1998) and
Bland-Hawthorn (this volume), it is possible to use the Stream
detections to determine what fraction of the ionizing photons produced
within the Galactic disk escape; this fraction $\fesc\approx
6\%$. This percentage is in reasonable agreement with the ionizing
photon flux needed to explain the diffuse ionized gas within the
Galactic disk (the Reynolds layer), and with models of photon
transport and escape within the Galaxy (Reynolds 1990; Miller \& Cox
1993; Dove \& Shull 1994a).

With the normalization provided by the Magellanic Stream detections,
it is possible to calculate the shape and intensity of the ionizing
radiation field produced by the Milky Way, and, in particular, to
calculate the expected emission measures for clouds near to the
Galactic disk. Ionizing photons escaping from the Galaxy will dominate
over the cosmic ionizing background out to a radius
\be
r_{\scriptscriptstyle MW}\sim 120\left({\fesc\over 0.05}{10^4\over
\phi_{\rm i,cos}}\right)^{1/2}\;{\rm kpc.}
\ee
The predicted emission measures can be used to determine distances for
the high-velocity clouds (Bland-Hawthorn \etal 1998). This method is
probably only reliable for heights $Z\gta 10$ kpc or so from the disk
midplane, so that structure in both the distribution of ionizing
photons and obscuring material within the disk is averaged out.
However, we note that the model presented in Bland-Hawthorn and
Maloney (1997, 1998) and Bland-Hawthorn \etal (1998) predicts the
distances to the M and A HVC complexes, detected in H$\alpha$ by
Tufte, Reynolds \& Haffner (1998), to within a factor of two. Future
H$\alpha$ observations at current levels of sensitivity will
revolutionize the study of HVCs.

\subsection{A Local Group Corona}
Is there a significant mass of gas associated with the Local Group?
Only a small fraction of the mass of baryons predicted by standard Big
Bang nucleosynthesis (\eg Olive, Steigman, \& Walker 1991) is actually
observed at $z=0$ (\eg Persic \& Salucci 1992). A recent inventory of
the observed baryons is given in Fukugita, Hogan, \& Peebles
(1998). If we parameterize the distribution of gas in a group (assumed
spherical) by the density distribution 
\be 
n(r)={n_o\over 1+(r/r_o)^2}
\ee 
where $n_o$ and $r_o$ are the core density and core radius,
respectively, then X-ray studies of poor groups of galaxies (which are
usually still substantially richer than the Local Group) typically
find the product 
\be 
n_o r_o T_{\rm keV}\sim 3\times 10^{20}\;\psqcm
\ee 
where \Tkev\ is the gas temperature expressed in keV. However, there
is a systematic dependence of the detection fraction on the morphology
of the group galaxies: groups dominated by ellipticals are generally
described by equation (8), with $\Tkev\sim 1$, while spiral-rich
groups generally do not show evidence for a diffuse intragroup medium;
only individual galaxies are detected (Pildis \& McGaugh 1996;
Mulchaey \etal 1996). The simplest interpretation is, of course, that
spiral-dominated groups do not have such an intragroup medium.
However, Mulchaey \etal (1996) suggested that the non-detections might
be due not to an absence of gas, but to lower gas temperatures in the
spiral-rich groups: the velocity dispersions of spiral-rich groups are
significantly lower than those of elliptical-rich groups, and imply
gas temperatures $\Tkev\approx 0.2-0.3$. Gas at such temperatures is
extremely difficult to observe directly, even at soft X-ray
wavelengths. In fact, from deep ROSAT images, Wang \& McCray (1993)
found evidence for a diffuse thermal component with T$_{\rm keV}$
$\sim$ 0.2 and electron density, $n_e \sim 1\times 10^{-2}\ x_{\rm
kpc}^{-0.5}$ cm$^{-3}$ (assuming primordial gas) where $x_{\rm kpc}$
is the line-of-sight depth within the emitting gas in
kiloparsecs. With no information at present available on the spatial
extent of the emission, it is impossible to say whether this is very
local gas (within the disk of the Galaxy, or a Galactic corona), or
more extended emission associated with the potential of the Local
Group itself. In the continuing infall scenario advocated by Blitz
\etal (1998) for the origin of the HVCs, a substantial fraction of the
infalling clouds will collide (preferentially near the barycenter of
the Local Group, approximately midway between M31 and the Milky Way)
and shock up to the virial temperature, $T_{\rm vir}\sim 0.2$ keV,
leading eventually to the formation of a group corona.

In Maloney \& Bland-Hawthorn (1998), we considered whether such a
Local Group corona could be detected indirectly, through the ionizing
photon flux produced as the gas cools. This can in principle be
important, for example in clusters of galaxies, where the resulting
flux can exceed the cosmic background by an order of magnitude.
However, in the Local Group, the other constraints which can be
imposed on such a corona -- in particular, dispersion measure
observations toward pulsars in the LMC and distant globular clusters,
the ``timing mass'', and the X-ray observations mentioned above --
rule out an ionizing flux from a corona which is significant (\ie
exceeds the probable cosmic background by a substantial factor) for
scales larger than a few tens of kpc. Basically, if the core radius is
large (so that the flux from the corona is important on the scale of
the Local Group), then the core density must be low to avoid violating
these other constraints. Such a corona could still contain a
substantial amount of mass: the direct observational limits on warm
gas in the Local Group are still not very stringent.

\subsection{Decay Photons from the Galactic Halo}
The immense number (of order $10^{76}$) of neutrinos in the halo of
the Milky Way in Sciama's model will produce a local ionizing photon
flux, as they slowly decay on a timescale $T_\nu\sim 10^{23}$
seconds. Assuming a nonsingular isothermal sphere for the neutrino
density distribution (as in equation [7]), the flux from a neutrino
halo can be written
\be
\phi_i \sim 2\times 10^6 \left({T_\nu\over 10^{23}}\right)^{-1}
{V_{100}^2\over (r_o^2 + r^2)^{1/2}}\;\phoflux
\ee
where $V_{100}$ is the (asymptotic) velocity of the halo scaled to
$100\kms$, and we have assumed that the galactic disk is completely
opaque to the decay photons, so that only one-half of the halo
contributes to the flux on either side of the disk. For the parameters
of Sciama (1998) and halo parameters appropriate to the Milky Way, the
predicted flux is $\phi_i\approx 2-3\times 10^5$ \phoflux\ at
$r\approx 10$ kpc (assuming a core radius of several kpc and a
rotation velocity of $200\kms$). This is in the range of the ionizing
photon fluxes found by Tufte \etal (1998) for HVC complexes A and
C. However, neither the Milky Way nor the LMC can produce the fluxes
necessary for the Stream detections through decay of halo neutrinos.

\section{Summary}
Throughout most of the Local Group volume, the ionization will be
dominated by the cosmic background, at a level $\phi\sim 10^4$
\phoflux\ which is still somewhat uncertain but which will be pinned
down by H$\alpha$ observations in the very near future. The cosmic
background itself is probably dominated by stellar photons escaping
from galaxies, if the escape fraction of photons from the Milky Way
(as determined from the detection of the Magellanic Stream in
H$\alpha$) is typical of $L_*$ galaxies. Within $r\sim 100$ kpc of the
Milky Way (depending on the angle with respect to the axis of the
Galaxy: see Bland-Hawthorn, this volume), the flux of ionizing photons
escaping the disk will exceed the background, but this is only of
order $1\%$ of the Local Group volume; a warm Local Group corona can
only influence a similarly small fraction of the LG volume. This has
very important implications for the suggestion of Blitz \etal (1998)
that the HVCs are relics of the formation of the Local Group (see also
Blitz, this volume): if most HVCs are actually at distances of order 1
Mpc, then they will be very faint in H$\alpha$ emission, as they will
be illuminated only by the cosmic background. This provides a direct
observational test of the model.

\acknowledgments
PRM acknowledges support from the Astrophysical Theory Program
through NASA grant NAG5-4061, and would like to thank JBH \& Sue for
their splendid hospitality. The authors sincerely hope that Dennis
will forgive us our conclusions regarding the decaying neutrino
model! Our deep admiration for him remains undiminished.


\end{document}